\documentclass[revtex4-1,preprint,
superscriptaddress,letter]{revtex4-1} 
\usepackage{graphicx}
\usepackage{dcolumn}
\usepackage{bm,textcomp}
\usepackage{units}
\usepackage{subfigure}
\usepackage{braket}
\usepackage{isotope}
\usepackage{hyperref}
\begin{document}
\title{A far-off-resonance optical trap for a Ba$^+$ ion}
\author{Thomas Huber}
\author{Alexander Lambrecht}
\author{Julian Schmidt}
\author{Leon Karpa}
\author{Tobias Schaetz}
\email{tobias.schaetz@physik.uni-freiburg.de}
\affiliation{
Albert-Ludwigs-Universit\"at Freiburg, Physikalisches Institut, Hermann-Herder-Strasse 3, 79104 Freiburg, Germany
}%

\date{\today}
\begin{abstract}
Optical trapping and ions combine unique advantages of independently striving fields of research. Light fields can form versatile potential landscapes, such as optical lattices, for neutral and charged atoms, avoiding detrimental implications of established radiofrequency (rf) traps while mediating interaction via long range Coulomb forces, controlling and detecting motional and electronic states on the quantum level.
Here we show optical trapping of $^{138}$Ba$^+$ ions in the absence of rf fields in a far-detuned dipole trap, suppressing photon scattering by three and the related recoil heating by four orders of magnitude. To enhance the prospects for optical as well as hybrid traps, we demonstrate a novel method for stray electric field compensation to a level below 9 mV/m.
Our results will be relevant, for example, for ion-atom ensembles, to enable four to five orders of magnitude lower common temperatures, accessing the regime of ultracold interaction and chemistry, where quantum effects are predicted to dominate.
\end{abstract}
\maketitle
Trapping of ions was demonstrated decades earlier than trapping
of neutral atoms albeit the concepts are closely related
\cite{Paul1989,Phillips1998}. A prominent reason is the much higher
well depth provided by radiofrequency (rf) traps, on the order of
$10^4$ K, compared to traps for neutral particles. Optical dipole
traps, for example, typically provide depths in the $10^{-3}$ K
range. The discrepancy is mainly due to the comparatively large
Coulomb forces that can be exerted on charged particles.\\
Over the course of the years, independent research on trapped ions
and atoms has yielded a remarkable growth of knowledge \citep{Wineland2012,Haroche2012} advancing various
topics, for example, quantum information \citep{Wineland2011,Schindler2011} and quantum
metrology \citep{Chou2010,Bloom2014,Rosenband2008,Huntemann2012,Madej2012}, or cold collisions \citep{Ospelkaus2010}, related to
chemistry and astrophysics. The advantages related to the charge of the ions, such as the unique controllability of electronic and motional degrees of freedom on the
quantum level and their characteristics to mediate long range
interactions, come at the price of their high sensitivity to stray electric fields. In rf traps, these fields displace the ion from the
node of the rf potential and result in an rf driven oscillation, the so called micromotion \citep{Paul1989}. Several techniques
have been developed to compensate stray electric fields
\citep{Berkeland1998,Barrett2003,Raab2000,Hoeffges1997,Ibaraki2011,Narayanan2011,Chuah2013}.
Nevertheless, residual rf micromotion still limits the progress in
several fields of research, such as metrology, where the
related second order Doppler shift contributes significantly to the
inaccuracy of atomic clocks \citep{Chou2010,Rosenband2008}.
Further improving the
compensation is crucial.\\
However, a fundamental impact of the rf field, even in the absence of any
stray electric field, has been revealed recently \citep{Cetina2012}
for the intriguing experiments carried out in hybrid traps
\citep{Smith2005,Grier2009,Zipkes2010,Schmid2010,Ravi2012,Harter2014}, where trapped neutral atoms or
molecules are spatially overlapped with an ion or ions confined in an rf
trap. Even if the ion is additionally considered as a classical
point-like particle at zero temperature at the minimum of the rf
potential, the polarization induced in an approaching atom in turn
leads to a displacement of the ion. This results in the rf field acting as a driving perturbation in the ion-atom ensemble, effectively pumping energy into the system. It has been shown that the common equilibrium temperature is not determined by the cold atomic bath (for Rb approximately 100 nK), but limited by the rf drive to the millikelvin regime \citep{Cetina2012}.\\
Replacing rf fields and confining ions optically is predicted to be advantageous in
several applications: to open up novel prospects in extended
potential landscapes, such as optical lattices, for arrays of ions \citep{Cirac00} or for exploiting the advantages of
trapping atoms and ions simultaneously
\citep{Schneider2012a,Cirac00,Doerk2010,Schmied2008,Kollath2007}, and to circumvent rf micromotion by
omitting any rf fields \citep{Cetina2012}.\\
We will emphasize the latter as one showcase and consider its
potential impact on the field of ultracold chemistry. Here, embedding ions
optically in quantum degenerate gases is predicted to allow reaching
temperatures 4-5 orders of magnitude below the current state-of-the-art \citep{Cetina2012}. In  this temperature regime quantum effects are expected to dominate, in (i) many-body
physics, including the potential formation and dynamics of
mesoscopic clusters of atoms of a Bose-Einstein condensate (BEC), binding
to the ``impurity ion'' \cite{Cote2002}, as well as in (ii)
two-particle s-wave collisions, the ultimate limit in ultracold
chemistry \citep{Ospelkaus2010,Krych2011,Cetina2012}.\\
Recently, optical trapping of a single ion has been demonstrated,
first in a single-beam dipole trap, superimposed by a static
electric potential \cite{Schneider2010,Schneider2012}. Subsequently,
all-optical trapping of an ion in an optical lattice (1D-standing
wave) has been achieved \citep{Enderlein2012}. In these
proof-of-principle experiments, the frequency of the dipole laser
has been tuned close to the relevant transition, severely
compromising the performance of the trap by the off-resonant
scattering of photons. \\
Additionally, optical confinement in one dimension has been
reported, pinning an ion or suppressing ion transport in a hybrid trap. Radially,  the
ion(s) remain confined  along two axes by an active linear rf trap,
while the third is overlapped with a far off-resonant
\citep{Linnet2012} and near-resonant, blue detuned
\citep{Karpa2013}, standing wave of an optical cavity,
respectively.\\
Here we show optical trapping of a $^{138}$Ba$^+$ ion in the absence of rf fields in a
far-off-resonance optical trap (FORT), suppressing photon scattering by three and the related recoil heating by four orders of magnitude. To further enhance the prospects of experiments in optical as well as in hybrid traps, we propose and implement a
generic method for sensing and compensating stray electric fields, currently to a level below 9 mV/m.
\section*{Results}
\subsection*{Optical trapping of Ba$^+$ in a far-off-resonance dipole trap}
\begin{figure}[h!]
     \centering
        \subfigure{
           \label{fig:TrapSeqA}
           \includegraphics[width =0.475 \textwidth]{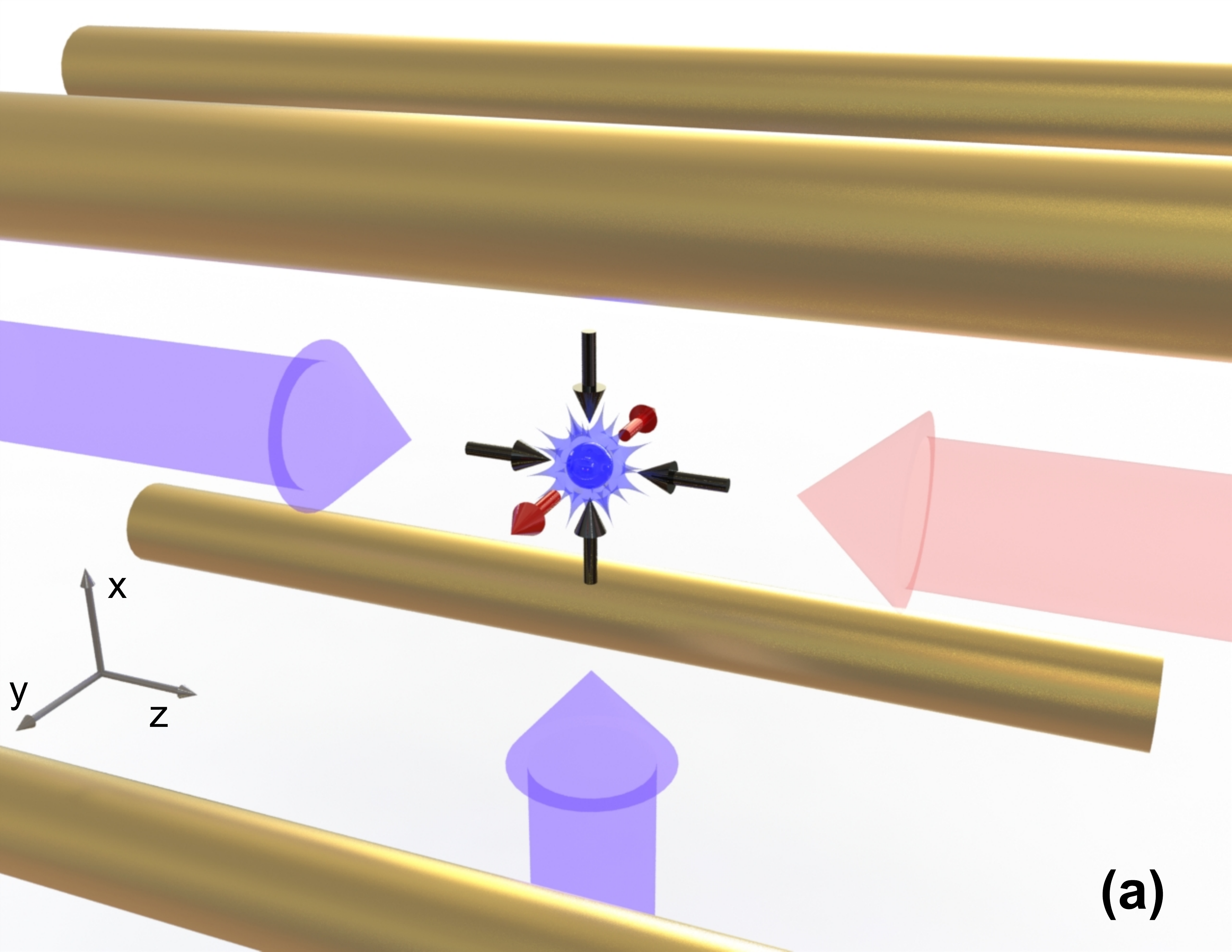}}
        \centering
     \subfigure{
           \label{fig:TrapSeqB}
           \includegraphics[width =0.475 \textwidth]{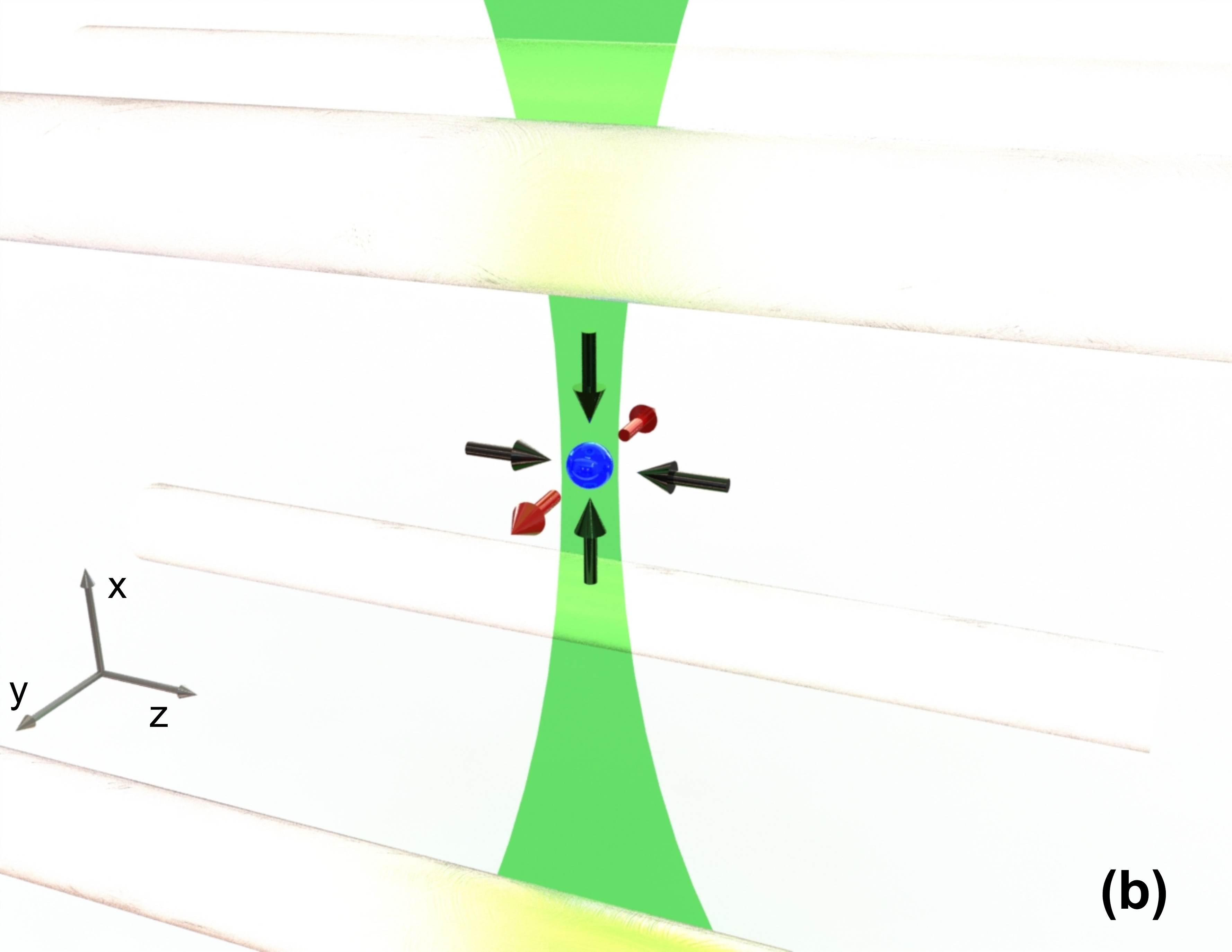}}
     \caption{Schematic of the setup (not to scale) and the optical ion trapping sequence. \textbf{(a)} We start by Doppler cooling the ion from two directions (blue arrows, from left and bottom) and repumping (red arrow, from right) from one direction in a linear rf trap (depicted as four quadrupole electrodes). After carrying out a coarse compensation of stray electric fields (see methodology), we turn off the cooling lasers and the repumping laser to assure pumping into the electronic ground state $\ket{S_{1/2}}$. The dc potential is chosen to be defocusing along the y axis, but focusing along the z and x axes, symbolized by the inwards (black) and outwards (red) pointing arrows, respectively. The distorted shape surrounding the ion symbolizes the driven rf micromotion, superimposed on the harmonic motion within the time-averaged trapping potential. \textbf{(b)} We turn on the dipole trapping laser (green) and switch off the rf trap (the deactivated rf drive is indicated by the translucent electrodes). That is, we provide confinement by optical dipole and dc potentials only. After a duration $\Delta t$, the rf trap, the cooling and the repumping laser are turned on again while the dipole trapping laser is turned off. In the case of successful optical trapping the ion gets finally detected with a CCD camera via resonant fluorescence in the rf trap again (see (a)).}
       \label{fig:sequtrap}
\end{figure}
The protocol for optical trapping is illustrated in Fig.~\ref{fig:sequtrap}. We still exploit an rf trap~\cite{Leschhorn2012,Kahra2012} (depicted in Fig.~\ref{fig:TrapSeqA}) to 
load \citep{Leschhorn2012b}, cool and finally detect the ion. In addition, we use the rf field to trap an ion for sensing and compensating stray electric fields and to optimize the spatial overlap of the rf and the optical trap, the latter 
being substantial for an efficient transfer. We start the optical ion trapping sequence by switching off the cooling beam (transition between the electronic states $\ket{S_{1/2}}\leftrightarrow \ket{P_{1/2}}$ at \unit[493]{nm}) and subsequently the repumping beam ($\ket{D_{3/2}} \leftrightarrow  \ket{P_{1/2}}$ at \unit[650]{nm}) to prepare the ion in the electronic ground state $\ket{S_{1/2}}$. We increase the power of the dipole trapping beam (at \unit[532]{nm}) linearly to its level $P$. After turning off the voltage supply of the rf trap, we electronically disconnect it. In the subsequent time span $\Delta t$ the ion is trapped by optical means assisted by carefully chosen and oriented dc potentials. After turning on the rf trap and turning off the dipole trap we employ resonant fluorescence detection to reveal if the optical trapping attempt has been successful.\\
\begin{figure}[t]
        \includegraphics[width =0.5 \textwidth]{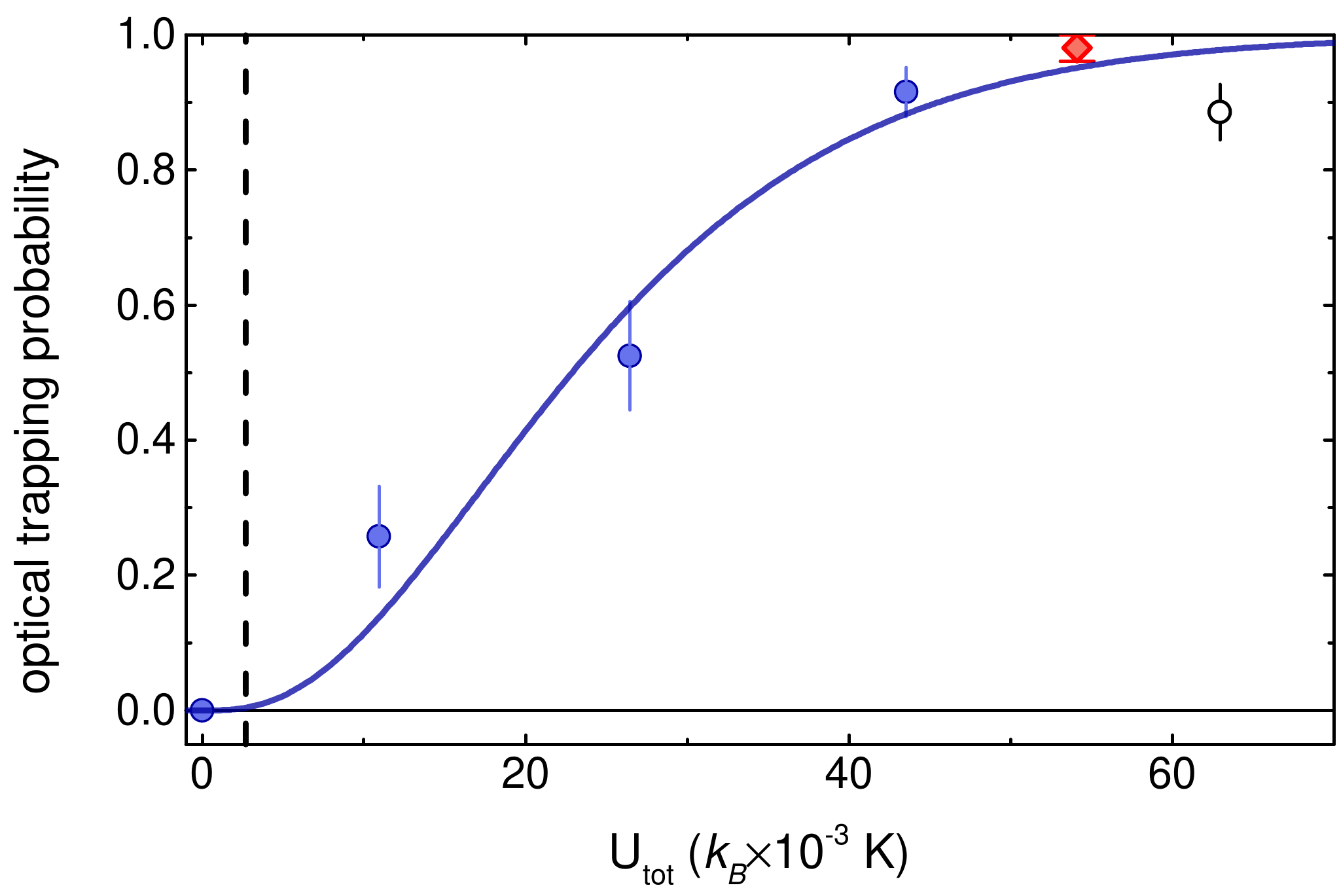}
    \caption{ 
    Optical trapping probability in dependence on the total trap depth $U_{tot}$, controlled by the power of the dipole laser beam. For each $U_{tot}$ we prepare the ions in the rf trap (see Fig.\ref{fig:TrapSeqA}), compensate stray electric fields and run a performance test at maximal optical trap depth, $U_{max} \approx k_B \times \unit[62]{mK}$ (black open circle). Subsequently, we reduce $U_{tot}$ to one of four values (blue filled circles), here for a trapping duration sufficiently short ($\Delta t = \unit[100]{\mu s}$) to avoid shelving into the $\ket{D}$ manifolds. To demonstrate optical trapping we compensate the stray electric field to a level of approximately \unit[80]{mV/m}, only. The corresponding electric force results in a threshold for the critical optical potential $U_{crit}$ depicted as the vertical dashed line. Applying our novel compensation method to its current limitation of \unit[8.7]{mV/m} (see text) will allow us to reduce the laser power of the dipole beam substantially.}       \label{fig:Prop2Pow}
\end{figure}
To derive the dependency of the optical trapping probability on the depth of the total potential $U_{tot}$ (optical and dc confinement), we first confirm that the cooling, repumping, stray field compensation and beam alignment are set appropriately by performing one test at the maximum trap depth $U_{max} \approx k_B \times \unit[62]{mK}$, $k_B$ denoting the Boltzmann constant, achieved with a trapping laser power of $P_{max} \approx \unit[13.1]{W}$ at a beam waist $W=3.9~\mu$m ($1/e^2$ of the intensity $I$). The average of all performance tests amounts to $(89 \pm 4) ~ \%$ and is depicted in Fig.~\ref{fig:Prop2Pow}. Subsequently, we decrease the power of the dipole laser to $P$, corresponding to one specific trap depth $U_{tot}$, and repeat the trapping sequence to determine the optical trapping probability $P_{trap}$ with a statistical uncertainty of less than $10~\%$ (1$\sigma$). The results for one specific set of parameters are shown in Fig.~\ref{fig:Prop2Pow} (blue filled circles). For $P = \unit[0]{W}$ $(U_{tot} = k_B\times\unit[0]{K})$, $P_{trap}$ is zero, demonstrating that residual trapping fields remain negligible and confinement in 3D for $P > \unit[0]{W}$ is achieved by means of optical dipole forces. We fit the measured data with a model function and derive an initial ion temperature of $T_{init} = \unit[(8.5\pm 0.7)]{mK}$ (see methodology).\\
For optimized conditions, e.g. optimizing the initial laser cooling at $U_{tot} = k_B\times\unit[55]{mK}$, we currently reach optical trapping probabilities 
up to $\unit[(98\pm 2)]{\%}$ (red diamond). The corresponding harmonically approximated oscillation frequency amounts to $\omega_{y} \approx \omega_{z} \approx \unit[2\pi\times 130]{kHz}$. As a lower bound for the average time between two off-resonant events of photon absorption out of the dipole beam ~\cite{Grimm2000}, we estimate a coherence time of $t_{coh} = \unit[0.7]{ms}$. We experimentally confirm the scattering rate $1/t_{coh} \approx \unit[1.4 \times 10^{3}]{s^{-1}}$ by measuring the rate of electron shelving into the manifolds of the $\ket{D_{5/2}}$ level, where the branching ratio for the decay from the $\ket{P_{3/2}}$ into the $\ket{S_{1/2}}$ and the $\ket{D_{5/2}}$ states amounts to 3:1. Two photon recoils of one absorption and emission event cause an energy transfer on the level of $k_B\times\unit[0.6]{\mu K}$. In comparison to former results on optically trapping Mg$^+$ ions reported in Ref.~\cite{Schneider2010}, with a related scattering rate of $\unit[7.5 \times 10^5] {s^{-1}}$ and a recoil energy of $ k_B\times \unit[10]{\mu K}$, we achieve a suppression of the recoil heating rate by four orders of magnitude.\\
\subsection*{Compensation of stray electric fields assisted by the ac Stark shift}
\begin{figure}
    \centering
        \includegraphics[width =0.65 \textwidth]{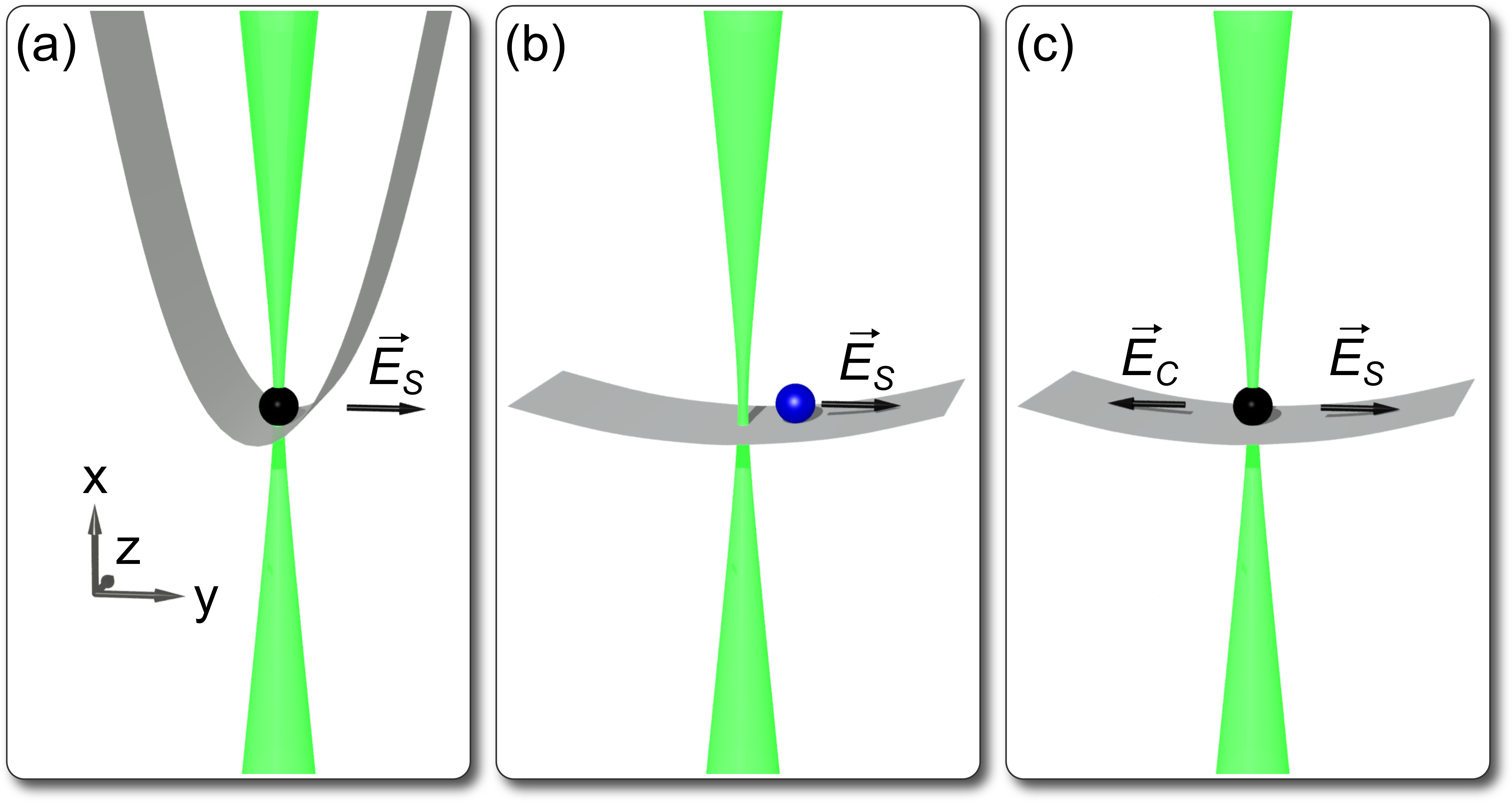}
    \caption{Schematic of the procedure to compensate a stray electric field $\vec{E}_S$ exploiting the ac Stark shift induced by a tightly focused dipole laser beam (green) as well as the related modulation of the fluorescence rate of the ion exposed to a detection beam (not shown), tuned close to the unshifted transition and of nearly constant intensity on the scale of relevant displacements. \textbf{(a)} The dipole beam is first characterized and centered on the ion (black sphere) at maximal confinement ($\omega_y = 2 \pi \times \unit[ (307 \pm 0.1) ]{kHz}$). \textbf{(b)} The confinement is switched to its minimal value ($\omega_y = 2 \pi \times \unit[( 27.3\pm 0.1 )]{kHz}$) and the ion (blue sphere) shifts into the new equilibrium position at lower intensity of the dipole laser beam and, thus, with its transition closer to resonance with the detection beam leading to increased fluorescence. \textbf{(c)} The ion is moved to its former position in (a) with the compensation field $\vec{E}_{C}$ induced by potentials applied to compensation electrodes along the y axis.}
      \label{fig:compensation}
\end{figure}

\begin{figure}
\centering
\includegraphics[width =0.5 \textwidth]{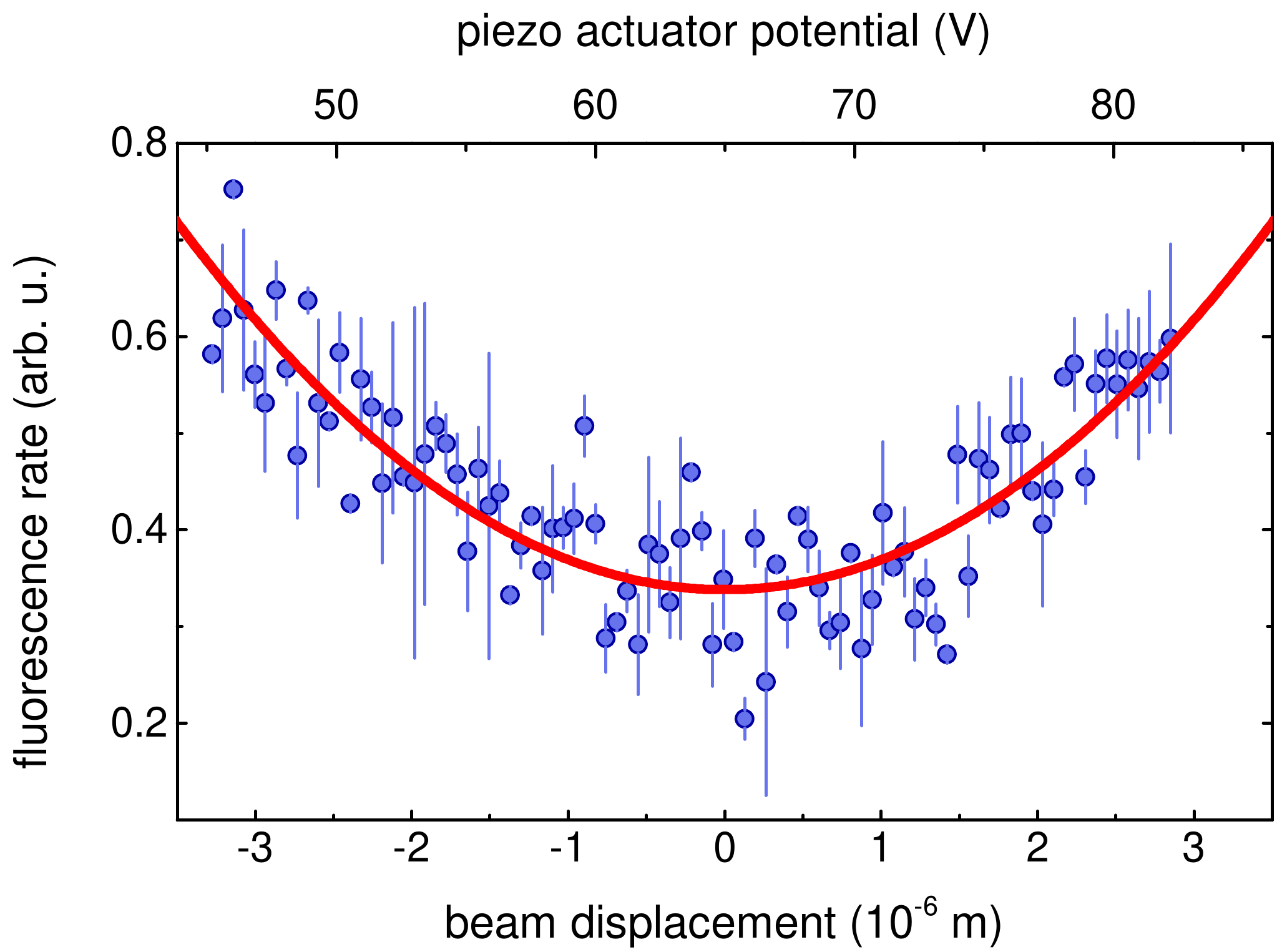}
\caption{
Fluorescence rate of an ion trapped in an rf field providing maximal confinement ($\omega_y = 2 \pi \times \unit[ (307 \pm 0.1) ]{kHz}$) in dependence on the potential applied to the piezo actuators and the displacement of the center of the dipole beam along the y direction, respectively. The dipole beam is inducing an ac Stark shift on the electronic energy levels of the ion, increasing the detuning of the cooling laser from the atomic transition and, thus, decreasing the fluorescence rate. The dependence of fluorescence in the vicinity of the center of the (Gaussian shaped) dipole beam, is approximated quadratically and its center determined by a least square fit (solid line) with an uncertainty of \unit[26]{nm}. The vertical error bars denote the statistical error after an integration time of 1 s per data point ($1\sigma$).
}
\label{fig:piezoscan}
\end{figure}

\begin{figure}[ht]
    \centering
        \includegraphics[width =0.5 \textwidth]{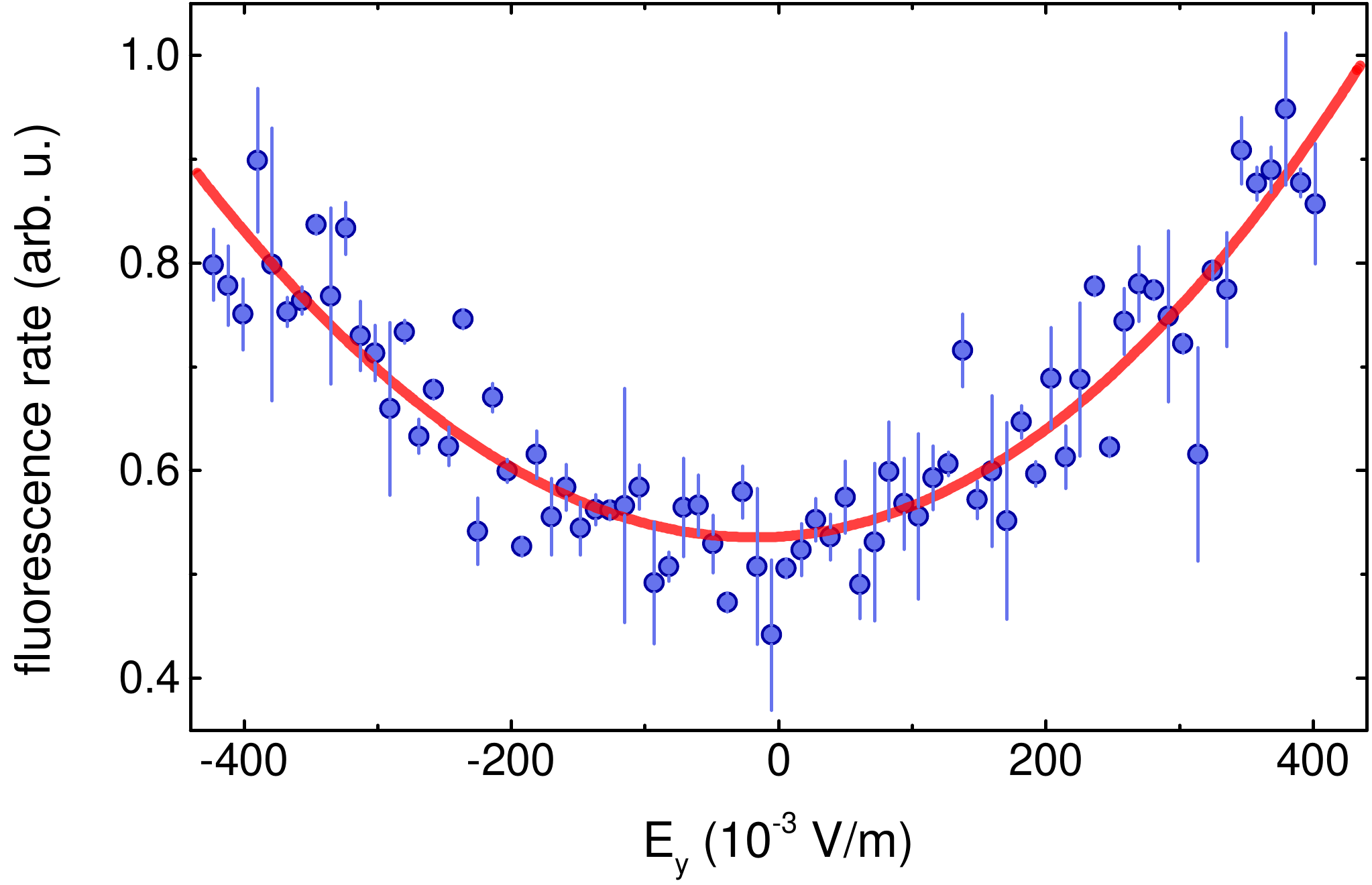}
    \caption{Fluorescence rate of an ion in dependence on the electric field caused by the potential applied to the trap electrodes along the y-direction. The ion probes the intensity profile of the dipole beam in the vicinity of its center. The center of this beam profile can be approximated by a least square fit of a quadratic function with an uncertainty of \unit[6.0]{mV/m} (solid line). The potential at the electrodes can be calibrated to an electric field, as described in the methods section, with a relative error of $\unit[11]{\%}$. The vertical error bars denote the statistical error ($1\sigma$) of the fluorescence rate.}
        \label{fig:efeld scan}
\end{figure}

To improve optical trapping of ions by further reducing the off-resonant scattering and improving coherence times seems straightforward. Following the established protocol for neutral atoms \citep{Grimm2000}, this involves increasing the detuning and the power of the dipole laser while reducing the initial temperature of the ion, in a first step down to the Doppler cooling limit at $T_D = \unit[300]{\mu K}$. However, in contrast to optically trapped neutrals, ions feature the additional charge directly coupling to electric fields, making them prone to stray electric fields and forces \citep{Cormick2011}. We have to overcome these by optical means, that is, to counteract by the gradient of the optical potential ($\propto \vec{\nabla} I$). Thus, our first objective is to further improve the sensitivity to detect and the capability to compensate stray electric fields.\\
We present a novel method for stray electric field compensation, applicable for many ion species and any trapping potential, demonstrated for a Ba$^+$ ion confined by the electric fields of a linear Paul trap. For this purpose we employ our tightly focused, attenuated dipole beam ($P \approx \unit[1]{W}$) to probe the position of the ion. The intensity profile causes a spatial dependency on the ac Stark shift of the ion's electronic levels, on the related detuning of our detection laser and therefore on the fluorescence rate.\\
Ideally, even a significant change of the rf confinement should not affect the position of the minimum of the rf potential. However, because the ion is exposed to a stray electric field $\vec{E}_S$, it is displaced to the minimum of the total potential, here provided by the superposition of rf and dc fields. Consequently, switching between two extremal rf amplitudes - that is, between the maximal and minimal secular frequencies that are still providing stable confinement - leads to two distinct locations of the ion probing different ac Stark shifts, with the related change of fluorescence providing an observable to derive $\vec{E}_S$
.\\
To apply our method, we (cool) detect the fluorescence of an ion in the rf trap by driving the cooling and repumping transitions while performing the following protocol (see Fig.~\ref{fig:compensation}):
\begin{enumerate}
\item \label{compzero} We employ a pre-compensation procedure consisting of two steps minimizing the fluorescence rate directly on the CCD camera. Firstly, at maximal rf confinement (currently at $\omega_y = 2 \pi \times \unit[(307\pm 0.1)]{kHz}$), we coarsely center the dipole beam on the ion (using micrometer screws on the objective mount), and secondly, minimize the ion's displacement at minimal rf confinement (currently  $\omega_y = 2 \pi \times \unit[ (27.3\pm 0.1) ]{kHz}$) by applying dc potentials to the compensation electrodes.
\item \label{compone}  We derive the profile at the center of the dipole beam and overlap it with the ion at maximal rf confinement. That is, we probe the fluorescence in dependence of 91 piezo controlled positions of the attenuated dipole beam, reveal its profile via a quadratic fit (see Fig.~\ref{fig:piezoscan}) and center it on the ion with the piezo-actuators.
\item \label{comptwo} We measure the displacement of the ion along the y axis due to the change of the rf confinement to its minimum amplitude. That is, we probe the fluorescence in dependency of the dc potentials applied to the (compensation) electrodes of the trap (see Fig.~\ref{fig:efeld scan}) at the position of the dipole beam obtained in step \ref{compone}.
\item \label{compthree} We derive the displacement of the ion from the beam profile via a quadratic fit and adjust the compensation field, $\vec{E_{C}}$, by the appropriate dc potential to recenter the ion.
\end{enumerate}
Depending on the amplitude of the initial stray electric field $\vec{E}_S$, we have to consider its initial impact on the derivation of the center of the rf trap in step~\ref{compone}. It leads to a systematic error, positioning the ion at the center of the total potential. However, this error can be mitigated by iterating our protocol (steps \ref{compone} - \ref{compthree}). Currently we estimate an upper bound for the uncertainty of the residual electric field of $\Delta E_{S} \approx \unit[8.7]{mV/m}$. A more detailed treatment of the remaining systematic errors and the limitations is given in the discussion and methods section. To benchmark our current results we compare them to those achievable by state-of-the-art techniques in rf traps for different applications, featuring a minimal electric field uncertainty in the range of a few \unit[100]{mV/m} (e.g. \citep{Narayanan2011}), and with specific experimental arrangements \unit[20]{mV/m}~\cite{Haerter2013}. Due to our improvements, even in hybrid traps, the achievable compensation level is within the limits that are predicted to be sufficient to enter the s-wave regime - at least for the few selected ion-atom combinations not being practically excluded already, that is, restricted to the largest ion-atom mass ratios, such as Li and Yb$^+$~\cite{Cetina2012}.\\
The method is additionally advantageous, especially for optical traps for ions, because it optimizes the spatial overlap between the optical and the initializing rf trap as well. It therefore mitigates the gain in the potential energy of the ion caused by the not fully adiabatic transfer between the traps. In summary, it allows to further reduce the required critical intensity of the optical trapping beam by more than a factor of 10 compared to the compensation level used for the optical trapping experiments described above (see Fig. \ref{fig:Prop2Pow}), further prolonging the coherence time accordingly. To take full advantage of the decreased (critical) potential depth $U_{crit}$ we will improve laser cooling to reach the standard Doppler limit at $\unit[300]{\mu K}$ and optionally increase the repumping efficiency~\cite{Berkeland2002} to keep the ion in the $\ket{S}$ manifold.
\section*{Discussion}
To take our method to the limits set by state-of-the-art technology we propose the following improvements: (i) decreasing the position uncertainty by further reducing $W$ of the diagnostics beam (dipole beam at low intensity) or by implementing a standing wave with a $\lambda/2$-periodic intensity pattern~\cite{Guthohrlein2001} decreasing the uncertainty of the electric field $\Delta E_{S}$ by a factor of $2 W/\lambda \approx 16$, and (ii) increasing the reproducibility of the piezo-control of the beam position by a factor of 100 by state-of-the-art equipment, which should result in $\Delta E_{S}<\unit[1]{mV/m}$, otherwise, to the best of our knowledge, approached only by exploiting Rydberg atoms \cite{Osterwalder1999}, a method  incompatible with rf fields. At this level and for a setup suited for trapping ions, we expect another class of systematic effects to affect the precision of our method, such as thermal expansion of and drifting potentials at the electrodes (see methodology). However, we emphasize that statistical fluctuations, such as beam pointing of the dipole laser or generic vibrations in the setup, lead to a symmetric broadening of the fluorescence signal and a reduced sensitivity, but do not result in systematic errors. To address all spatial degrees of freedom a second dipole laser beam is advantageous. For our current application of choice, this can be achieved, for example, by exploiting a crossed beam arrangement common in all-optical BEC experiments~\citep{Arnold2011}.\\
To put the current results of the suppressed photon scattering and stray field uncertainty in a relevant perspective we consider the prospects for experiments of optically trapped ions immersed in a bath of ultracold atoms. To achieve efficient (sympathetic) cooling and to reach nanokelvin temperatures by elastic scattering of the ion with ultracold atoms~\citep{Krych2011}, we have to compare the residual recoil heating rate due to off-resonant photon scattering with the (in)elastic scattering rate between Ba$^+$ and Rb, responsible for the sympathetic cooling (chemical reaction). Already at our current level of stray field compensation and reduced power required in the dipole beam, on the order of 100 undisturbed elastic scattering events should take place ~\citep{Harter2014} before the first occurrence of a photon recoil event disrupting coherent processes and limiting sympathetic cooling. The rate for inelastic collisions is suppressed by another factor of approximately $10^{4}$~\citep{Krych2011}. We estimate that to allow for sympathetic cooling to the temperature of equilibrium with the atomic bath, a few $(<10)$ collisions are sufficient~\cite{metcalf1999laser}, thus recoil heating remains negligible. Consequently, the current realization is already sufficient to study ultracold interactions, e.g. the potential formation of mesoscopic clusters~\cite{Cote2002}, and to extend and alter the possibilities of studying long-range interaction in homonuclear systems of Rydberg atoms within BECs \citep{Balewski2013}. On the other hand, the inelastic (incoherent) processes, such as molecule formation in the quantum regime of s-wave scattering are of interest in their own right~\citep{Grier2009,Schmid2010,Zipkes2010,Krych2011,Ravi2012,Harter2014}.\\
In our current implementation, the trapping laser at \unit[532]{nm} remains blue detuned with respect to the transitions between the electronic $\ket{P}$ and the metastable $\ket{D}$ states of Ba$^+$. For our envisioned application, shelving into the sublevel manifolds of the $\ket{D}$ states is not desirable, since the related deconfinement leads to the loss of the ion and thus derogates statistics. There are several options to minimize or mitigate this effect. (i) A second, further detuned, laser (of a bichromatic trap) at \unit[1064]{nm} - initially mainly responsible for the optical trapping of Rb - can also be used to simultaneously trap Ba$^+$ (switching off the \unit[532]{nm} beam after sympathetic and evaporative cooling). At sufficiently low temperatures, such substantially increased detuning should make it possible to confine the ion in the $\ket{S}$ and $\ket{D}$ states simultaneously. (ii) Already implemented laser beams at \unit[615]{nm} and \unit[650]{nm} can be used to resonantly repump the population into the electronic ground state. Optionally, the dependence of the detuning due to the spatially modulated ac Stark shift can be mitigated by interchangeably exposing the ion to the dipole and the repumping beams at a rate high compared to the secular frequency within the optical potential. (iii) As an alternative approach, tuning the frequency of the dipole laser blue of resonance, in combination with confining dc fields, might allow us to further reduce off-resonant excitation while providing trapping in $\ket{S}$ and $\ket{D}$ manifolds simultaneously.
\section*{Methods}
\label{methods}
\subsection*{Initial coarse compensation of stray electric fields}
In order to mitigate the systematic error related to the initial displacement of the ion from the node of the rf potential in presence of the finite stray electric field, we employ the following routine (step (\ref{compzero})) before performing the compensation protocol steps (\ref{compone}) to (\ref{compthree}) (see main text). This preparatory procedure with a total duration of approximately 60 s has to be carried out approximately once a day. We note that iterating steps (\ref{compone}) to (\ref{compthree}) of the described scheme without the preparatory step (\ref{compzero}) allows us to compensate within a measurement duration of $n$ times 167 s, to a level of $\left| \Delta E_S\right| = \unit[8.7]{mV/m}$ or lower. This precision can also be reached after a single cycle if the initial stray electric field remains smaller than $\left|\Delta E_S^0\right| = \unit[140]{mV/m}$. To this end, it is sufficient to locate the rf node with an uncertainty of approximately $W_y /2$ at maximum rf confinement and subsequently position the ion within this distance at the dipole beam center at minimal rf confinement, confirmed by a readily detectable drop of the fluorescence rate to approximately 50 \%, respectively. The corresponding positioning errors with the chosen secular frequencies of $\omega_y = 2 \pi \times \unit[(307 \pm 0.1)] { \, kHz}$ and $\omega_y = 2 \pi \times\unit[(27.3\pm 0.1)]{ \, kHz}$ and a waist of $W_y = \unit[4]{\mu m}$ amount to $\left|\Delta E_S^0\right| \approx \unit[140]{mV/m}$.
\subsection*{Total uncertainty in the compensation of stray electric fields}
We assume that the electric field has already been compensated to the level of $\left|\Delta E_S^0\right| = \unit[140]{mV/m}$ as described above (see Fig.~\ref{fig:compensation}).
At maximal confinement ($\omega_y = 2 \pi \times \unit[(307 \pm 0.1)]{ \, kHz}$), $\left|\Delta E_S^0\right|$ results in a displacement of $d_{res} \approx \unit[26]{nm}$. In addition, the position of the center of the dipole beam is estimated as the standard error ($1\sigma$) of the least square fit (see Fig.~\ref{fig:piezoscan}) to $d_{fit} = \unit[56]{nm}$. The dipole laser beam is shifted to discrete locations with a piezo-actuated mirror that can be reproducibly positioned with an uncertainty of $d_{rep} = \unit[94]{nm}$ with a minimal incremental step size of $\unit[2]{nm}$. These uncertainties are added quadratically: $d_{align} = \sqrt{d_{res}^2+d_{fit}^2 + d_{rep}^2} \approx \unit[112]{nm}$. At the minimal confinement of $\omega_y = 2 \pi \times \unit[(27.3\pm 0.1)]{kHz}$, this yields an uncertainty of the residual electric field of $\Delta E_{align} \approx \unit[4.6]{mV/m}$.\\
The uncertainty of centering the ion in the dipole trapping beam via potentials on compensation electrodes is obtained from the standard error of the least square fit amounting to $\Delta E_{pos} \approx \unit[6.0]{mV/m}$ ($1\sigma$, see Fig.~\ref{fig:efeld scan}).\\
The pointing of the beam, controlled by the piezo-mounted mirror, is drifting systematically during the measurements, which take $\approx \unit[91]{s}$ for the laser alignment in step (\ref{compone}) and $ \approx \unit[76]{s}$ for the field compensation in step (\ref{compthree}). We measure a corresponding drift rate of \unit[0.5]{nm s$^{-1}$} and, without correcting for such systematic drifts after performing steps (\ref{compone}) and (\ref{compthree}), assume that they simply accumulate. These drifts then lead to $\Delta E_{sys1} \approx \unit[0.9]{mV/m}$ and $\Delta E_{sys2} \approx \unit[0.7]{mV/m}$, respectively. Thus, as a worst case scenario, we estimate the total electric field uncertainty $\Delta E_{S}$ by linearly adding statistic and systematic errors to $\Delta E_{S} = \sqrt{\left(\Delta E_{sys1}+\Delta E_{align}\right)^2+\left(\Delta E_{sys2}+\Delta E_{pos}\right)^2} \approx \unit[8.7]{mV/m}$.\\
The impact of other systematic effects has been mitigated considering the argument presented in the following. (i) The effect of light pressure of the cooling light on the ion can lead to significant displacement, and consequently to misalignment of the dipole beam. To minimize light forces along the y axis we currently provide efficient cooling along the z- and x-direction only. In the future we will implement pairs of counter-propagating beams to provide optimal cooling towards the Doppler limit while additionally eliminating the net light pressure in all degrees of freedom. (ii) Effects of dipole forces induced by the dipole beam are minimized by substantially attenuating the dipole trapping beam. (iii) The effects of thermal drifts of the trap geometry due to changes in rf amplitude are still negligible. Based on conservative estimates and measurements of the temperature we conclude that the rf electrodes do not warm up by more than $\unit[0.3]{K}$, corresponding to an upper bound for a spatial drift of the electrodes of \unit[10]{nm}.\\
\subsection*{Calibrating the electric field for position control and compensation}
We derive the amplitude of the electric field provided at the position of the ion by calibrating the potential applied to a compensation electrode to the ion displacement at a confinement of  $\omega_y = \unit[2\pi \times (307 \pm 0.1)]{kHz}$. We start by moving the dipole 2beam across the position of an ion in the rf trap. We verify that the induced ac Stark shift $\Delta_S$ is smaller than the natural linewidths of the $\ket{S_{1/2}} \leftrightarrow \ket{P_{1/2}}, \ket{P_{3/2}}$ transitions of the ion and derive the beam waist $W_y$ along the y axis. A displacement of the beam by $W_y$ in a harmonic potential with a secular frequency of $\omega_y$ requires a force $F_y = m \omega_y^2 W_y$, $m$ being the mass of the ion. It can be related to an electric field $E_C$ along y with $F_y = e E_C$, where $e$ is the elementary charge.\\
Subsequently, we keep the laser beam fixed and move the ion along the y-direction by applying a potential $V_y$ to the corresponding electrodes. The recorded fluorescence pattern can be fitted by a Gaussian beam intensity profile yielding its width in units of the applied potential and, thus, the electric field $E_C$. The calibration of the electric field to an applied potential is in agreement with numerical simulations using a finite difference method. The uncertainty of the residual stray field of $\Delta E_S = \unit[8.7]{mV/m}$ is equivalent to an electric potential uncertainty of $\unit[19]{\mu V}$ on two adjacent electrodes.
\subsection*{Optical trapping of Ba$^+$ in a far-off-resonance trap}
To derive the total trap depth $U_{tot}$ depicted in Fig.~\ref{fig:Prop2Pow}, we perform a Monte Carlo simulation where we consider the center of the dipole trap beam ($y_0$), its waist ($W_y$), the dc harmonic frequency $\omega_y$ and the residual stray electric field $\Delta E_{S}$ to be normally distributed. From $5 \times 10^4$ randomly generated values for the relevant variables we numerically calculate the total trap depth along the y-direction (the direction with a repulsive dc potential, see Fig.~\ref{fig:sequtrap}) by determining the difference between the minimum and the smallest local maximum of the trapping potential
\begin{equation}
U(y) = U_0 e^{-2 (y-y_0)^2/(W_y^2)} - \frac{1}{2} m \omega_y^2 y^2 - e \Delta E_{S} y\text{,}
\end{equation}
where $U_0$ is the depth of the optical trapping potential~\cite{Grimm2000}. By binning the values of $U_{tot}$ we obtain the probability distribution for the total trap depth and thus the mean value as depicted in Fig.~\ref{fig:Prop2Pow} as well as the corresponding standard deviation.\\
To obtain the temperature from the optical trapping probability (see Fig.\ref{fig:Prop2Pow}) for a given $U_{tot}$ we consider the radial cut-off model introduced in Ref.~\cite{Schneider2012}
\begin{equation}
P_{trap} = 1- e^{-2\xi}-2\xi e^{-\xi},
\end{equation}
with $\xi = U_{tot}/(k_B T)$ where we assume the recoil heating rate to be negligible for the duration of the optical trapping.\\
The optical confinement along the focal direction (x axis) of the dipole trap is weak compared to the confinement along the y and z axes. We provide additional dc confinement along the x axis, which we achieve by tilting the principle axis with an accuracy of $\pm \unit[25]{^{\circ}}$ with respect to the laboratory frame resulting in a bare dc confinement with $(\omega_{x}^{dc})^2 = (\unit[2\pi \times (9.7\pm 0.5)]{kHz})^2$,  $(\omega_{y}^{dc})^2 = -(\unit[2\pi \times (16.7\pm 0.5)]{kHz})^2$ and $(\omega_{z}^{dc})^2 = (\unit[2\pi \times (13.6\pm 0.1)]{kHz})^2$.
\section*{Acknowledgments}
We thank D. Leibfried for helpful discussions and M. Wyneken for reviewing the manuscript. L. K. gratefully acknowledges support by the Alexander von Humboldt Foundation.
\section*{Author contributions}
T. H. planned and carried out the experiment, contributed to evaluation and wrote the manuscript. All authors contributed equally to performing the experiment, evaluating the data and assisted in writing the manuscript. 
%
%
\bibliography{main}
\end{document}